\documentclass[
reprint,
superscriptaddress,
 amsmath,amssymb,aps,
prl,
]{revtex4-2}

\usepackage{graphicx}
\usepackage{dcolumn}
\usepackage{bm}
\usepackage{xcolor}
\usepackage[T1]{fontenc}

\begin{document}

\preprint{APS/123-QED}

\title{Two-dimensional superconductivity in a thick exfoliated kagome film}

\author{Fei Sun}
\email{Correspondence: Fei.Sun@cpfs.mpg.de}
\affiliation{Max Planck Institute for Chemical Physics of Solids, 01187 Dresden, Germany}

\author{Andrea Capa Salinas}
\affiliation{Materials Department, University of California Santa Barbara,
Santa Barbara, California 93106, USA}

\author{Stephen D. Wilson}
\affiliation{Materials Department, University of California Santa Barbara,
Santa Barbara, California 93106, USA}

\author{Haijing Zhang}
\email{Correspondence: Haijing.Zhang@cpfs.mpg.de}
\affiliation{Max Planck Institute for Chemical Physics of Solids, 01187 Dresden, Germany}

\date{\today}

\begin{abstract}
We report the observation of two-dimensional superconductivity (2D SC) in exfoliated kagome metal CsV$_3$Sb$_5$ with a thickness far thicker than the atomic limit. By examining the critical current and upper critical magnetic fields ($H_{c2}$) of 40-60 nm thick films in the superconducting state, we identify a pronounced Berezinskii-Kosterlitz-Thouless (BKT) transition behavior, i.e. a drastic decrease of the superfluid stiffness near the transition, and a cusp-like feature of the angular dependent $H_{c2}$, both of which serve as direct evidence of 2D SC. In addition, an exceeding of the Pauli paramagnetic limit of the in-plane $H_{c2}$ is consistent with the 2D SC nature. The observed 2D SC occurs in thick films with the highest superconducting transition temperature $T_c$ and the lowest charge density wave transition temperature $T_{\rm {CDW}}$, which suggests that the charge density wave states are interrelated with the superconducting states. Our findings impose constraints in understanding the enhancement of SC in kagome superconductors, and illuminate pathways for achieving novel 2D superconducting states in more stable and much thicker systems. 
\end{abstract}

\maketitle
Two-dimensional superconductivity (2D SC) has attracted significant research interest for nearly a century, beginning with the observation of superconductivity in metallic thin films \cite{shalnikov1938superconducting,saito2016highly}. In recent decades, benefiting from the technical advances in modern fabrication, there have been marked developments in 2D SC across various systems, including cuprates \cite{terashima1991superconductivity, dekker1992absence}, oxide heterostructures \cite{reyren2007superconducting, biscaras2013multiple}, disordered thin films \cite{fiory1983superconducting,postolova2015nonequilibrium,weitzel2023sharpness} and van der Waals 2D materials \cite{lu2015evidence, costanzo2016gate}. 
Substantial efforts are still invested in 2D SC nowadays due to its outstanding properties, such as Berezinskii-Kosterlitz-Thouless (BKT) transitions \cite{berezinskii1971destruction, kosterlitz1972long} and the violation of the Pauli paramagnetic limit \cite{lu2015evidence, cao2021pauli}. Moreover, highly crystalline 2D superconductors have also been proposed as candidates for high-temperature superconductors \cite{gozar2008high, wang2012interface}. However, besides the technical challenges in realizing the atomic 2D limit, 2D superconductors are often intrinsically fragile and subject to quality degradation, which greatly limits their potential applications and the understanding of its intrinsic properties. Therefore, the pursuit of achieving 2D SC in a more stable and easily accessible system has garnered significant attention. One notable example along this direction is the realization of 2D SC in a bulk superlattice consisting of 2H-NbS$_2$ and a commensurate block layer \cite{devarakonda2020clean}, which offered a new pathway to study 2D SC. Here, we report on the direct observation of 2D superconductivity, via measurements of inverse penetration depths, in exfoliated nano-flakes of CsV$_3$Sb$_5$ crystals with thickness far thicker than the atomic limit. 

\begin{figure}[!b]
\includegraphics[width=3.3in]{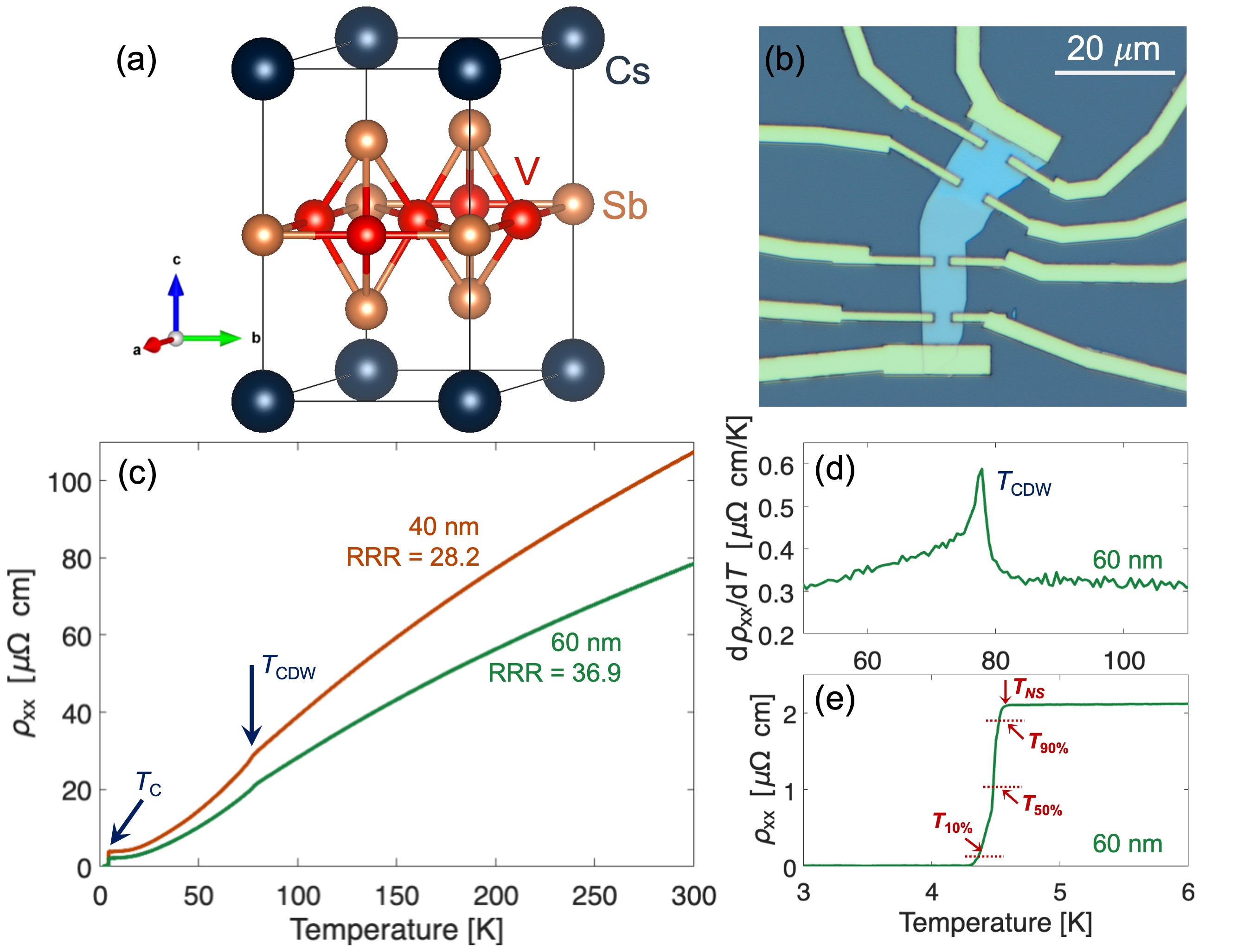}% 
\caption{\label{Fig1} (a) The lattice structure of CsV$_3$Sb$_5$ ($a$=$b$=5.5 \AA, $c$=9 \AA). (b) Optical image of a microfabricated device.  (c) $\rho_{xx}$ vs $T$ of 40- and 60-nm-thick samples. (d) and (e) The zoom-in plots of the CDW and the SC transition, respectively. The characteristic temperatures, determined by 90\%, 50\%, and 10\% of the normal-state resistance, are marked by arrows.}
\end{figure}

\begin{figure*}[ht]
\includegraphics[width=7in]{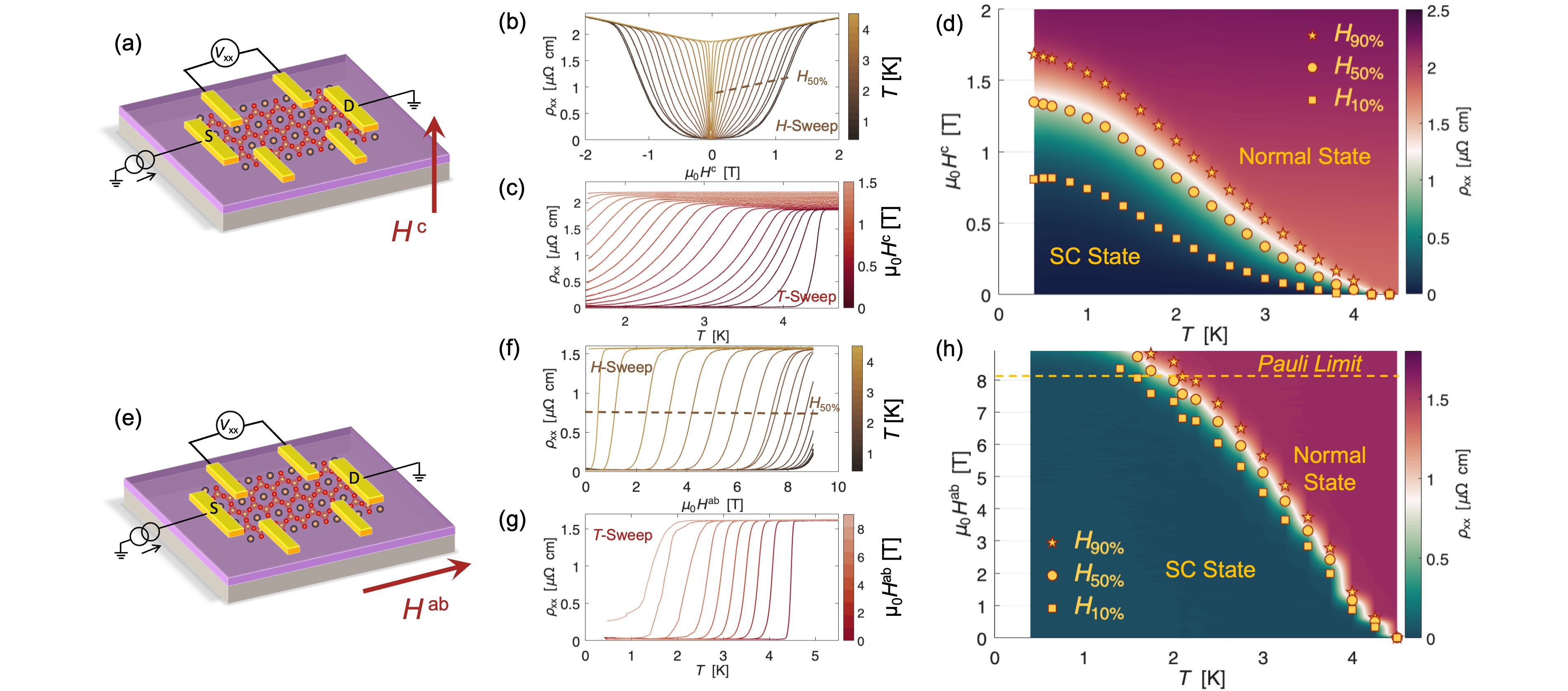}% 
\caption{\label{Fig2} The $H$-$T$ phase diagrams. (a)-(d) are under the out-of-plane field and (e)-(h) are under the in-plane field. (b) and (f) $\rho_{xx} $ vs $H$ measured at different $T$. (c) and (g) $\rho_{xx}$ vs $T$ measured at different $H$. (d) and (h) The $H$-$T$ phase diagrams with the magnetic field applied along different directions. }
\end{figure*}

CsV$_3$Sb$_5$ is a recently discovered kagome superconductor, which has emerged as a versatile platform for investigating the interplay of correlated electronic states, owing to its unique electronic structure and inherent frustration \cite{ortiz2020cs,xiang2021twofold,chen2021roton,nie2022charge}. Despite a number of theoretical predictions suggesting the potential existence of unconventional superconducting states at the Van Hove singularities \cite{PhysRevLett.127.177001, PhysRevLett.127.046401}, the majority of recent experimental observations indicate that the SC of bulk CsV$_3$Sb$_5$ is nodeless, with an anisotropic $s$-wave pairing \cite{duan2021nodeless,gupta2022microscopic,roppongi2023bulk}. While considerable efforts have been made to tune the SC properties via mechanical exfoliation \cite{song2021competition,song2023anomalous, zhang2023nodeless}, a non-monotonic variation of $T_{\text c}$ with the thickness reduction has been reported \cite{song2021competition,song2023anomalous}, which reveals an optimal thickness ranging from 40 to 60 nm with maximal $T_{\text c}$. The non-monotonicity categorizes the exfoliated flakes with optimal thickness as a special case of SC. The central question lies in understanding the role of dimensionallity in SC with thickness reduction, and thus, a thorough investigation of the SC ground state properties is highly desired.

In this work, we focus on the exfoliated nano-flakes of CsV$_3$Sb$_5$ with the optimal thickness. We extract the temperature dependent penetration depths from critical current and $H_{c2}$ measurements. We observe an abrupt drop of the superfluid stiffness, i.e. normalized inverse penetration depths, around the transition temperature, which is a hallmark of the BKT transition. A cusp-like feature of the angular dependent $H_{c2}$ is also observed, serving as direct evidence of the 2D SC. In addition, an exceeding of the Pauli paramagnetic limit of the in-plane $H_{c2}$ and a broadening of superconducting transitions under out-of-plane magnetic fields have been observed, providing further support for the 2D SC nature. The observation of 2D SC in the optimal thickness - nearly 70 times that of the interlayer spacing - indicates the existence of hidden correlations in this kagome superconductor.

The crystal structure of CsV$_3$Sb$_5$ is shown in Fig. 1(a). It crystallizes into the P6/mmm space group, with a kagome network of V and Sb layers separated by Cs layers. The single crystals used in this work are grown by self flux method \cite{ortiz2020cs}. For transport measurements, we realized microfabricated devices based on exfoliated crystals with thicknesses ranging from 40 to 300 nm. Fig. 1(b) is an optical image of a device as measured in this work (see Ref. \cite{Supplement} for fabrication details).

Figure 1(c) shows the temperature ($T$) dependence of in-plane resistivity $\rho_{xx}$ for both 40- and 60-nm-thick exfoliated crystals. All exfoliated crystals exhibit a metallic behavior. The residual resistance ratio (RRR) for the exfoliated crystals calculated using the room-temperature resistance and the resistance at 5 K varies from 20 to 40. Two characteristic temperatures are observed (Fig. 1(c)), corresponding to the transitions into the charge density wave state ($T_{\rm{CDW}}$) and the superconducting state ($T_{\rm{c}}$), respectively, and the detailed view of the two transitions are shown in Figs. 1(d-e). In contrast to the bulk crystals ($T_{\rm{CDW}}$=94 K and $T_{\rm{c}}$=2.5 K \cite{ortiz2020cs}), the 60-nm-thick film exhibits a lower $T_{\rm{CDW}}$ at 77 K and a much higher $T_{\rm{c}}$ at 4.47 K (defined as 50\% of the normal state resistance, $R_{\rm{NS}}$). The superconducting transition, which sets in at around $T$=4.55 K and reaches zero resistance at $T$=4.3 K, is substantially sharper than those observed in bulk single crystals \cite{ortiz2020cs}.

To further explore the SC nature, we perform magneto-transport measurements on the 60-nm-thick sample, by applying both out-of-plane ($H^{c}$) and in-plane magnetic field ($H^{ab}$). As shown in Figs. 2(a-d), the SC transition is suppressed with a perpendicular magnetic field. The upper critical field, $H_{c2}^c(T)$, is defined as the magnetic field at which the resistance reaches 50\% of $R_{\rm{NS}}$. At temperatures around 0.75 $T_{\rm{c}}$, a concave feature in $H_{c2}^c(T)$ is seen, which is an indication of multi-gap superconductivity \cite{PhysRevB.66.180502}. Upon further cooling, $H_{c2}^c(T)$ saturates to approximately 1.36 T at the zero-temperature limit, corresponding to an in-plane coherence length $\xi_{ab}(0)$ = $\sqrt{\Phi_0/(2 \pi H_{c2}^c(0))}$= 15.4 nm, where $\Phi_0$ is the flux quantum. We note that prominent Shubnikov-de Haas (SdH) quantum oscillations are observed in this temperature range at magnetic fields higher than 3 T~\cite{Supplement}, indicating a good electronic quality in the exfoliated crystals.

In contrast to $H_{c2}^c(T)$, the in-plane upper critical field, $H_{c2}^{ab}(T)$ shows a steeper $T$-dependence near $T_{\rm{c}}$ (Fig. 2(h)). Significantly, $H_{c2}^{ab}$ exceeds the Pauli paramagnetic limit $H_{\rm{PL}}=1.84$ $T_c $ [T/K] \cite{chandrasekhar1962note,clogston1962upper} at $T$ < 2 K (0.45 $T_{\rm{c}}$). According to the Werthamer-Helfand-Hohenberg (WHH) model $H_{c2}(0) \approx 0.73$ $T_{\rm{c}}(-dH_{c2}/dT)|_{T_{\rm{c}}}$ describing the orbital limit \cite{Helfand1966temperature}, a zero-temperature $H_{c2}^{ab}(0)$ =11.9 T is extracted, which exceeds the Pauli limit by 45\%. The anisotropic ratio ($H_{c2}^{ab}/ H_{c2}^{c}$) varies from 40 to 9 with respect to temperature~\cite{Supplement}, indicating an anisotropic SC nature. Regardless of the microscopic origin of $H_{c2}$ exceeding the Pauli limit, the absence of violations of the Pauli limit in the bulk crystals \cite{ni2021anisotropic} suggests that the thickness reduction may play a role on the enhancement of $H_{c2}^{ab}$ in the exfoliated crystals. In checks made on a 40-nm-thick device, similar $H$-$T$ diagrams is observed without further enhancement of $H_{c2}^{ab}$ \cite{Supplement}. 

\begin{figure}[t]
\includegraphics[width=3.3in]{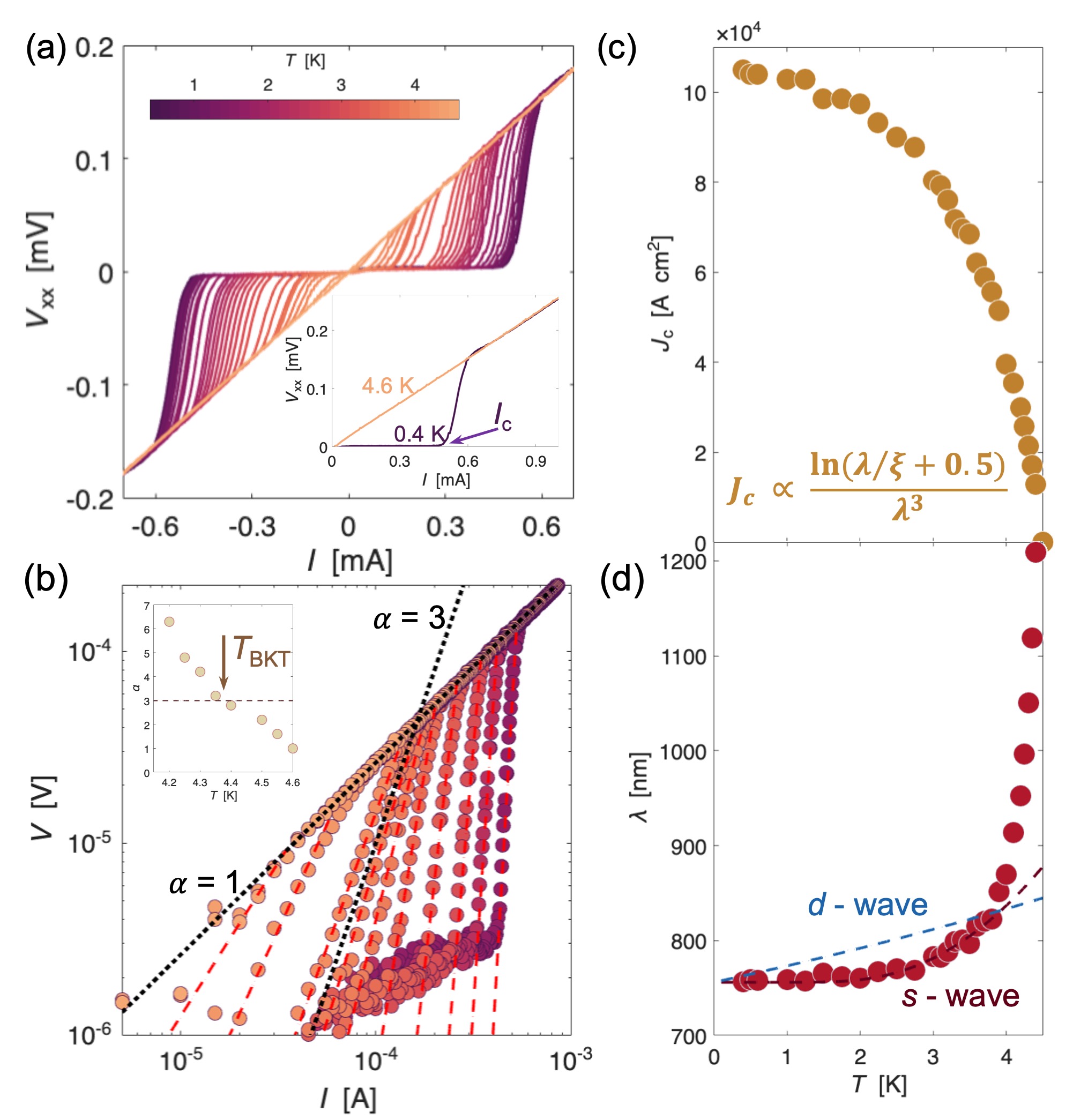}% 
\caption{\label{Fig3} (a) The temperature dependent $V$-$I$ curves. The inset shows how we define $I_c$. (b) The $V$-$I$ curves plotted on a logarithmic scale. Inset: Temperature dependence of $\alpha$ extracted from $V \propto I^\alpha$. (c) and (d) Temperature dependent critical current density $J_c$ and the calculated penetration depth $\lambda$. }
\end{figure}

To delve deeper into the exceeding of Pauli limit in the thin flakes, a thorough understanding of the superconducting gap and its pairing symmetry is crucial. Hence, we performed systematic critical current $I_{\rm{c}}$ investigations on the 60-nm sample. The voltage - current ($V$-$I$) studies enable analysis in the BKT scenario to assess the role of dimensionality in SC \cite{berezinskii1971destruction,kosterlitz2018ordering,costanzo2016gate,lu2015evidence}; meanwhile, combining $I_{\rm{c}}$  with the previously obtained coherence length $\xi$ facilitates a direct calculation of the magnetic penetration depth $\lambda$~\cite{talantsev2015universal}, whose $T$-dependence provides insights into the SC pairing symmetry. Figure 3(a) shows the $V$-$I$ curves for temperatures from 4.6 K to 0.4 K (from above $T_{\rm{c}}$ to 0.1 $T_{\rm{c}}$). A gradual crossover from a linear dependence at $ T>T_{\rm{c}}$ to a non-linear dependence at $ T<T_{\rm{c}}$ with a plateau at zero is observed as expected for superconducting transitions. Interestingly, from the $V$-$I$ curves plotted on a logarithmic scale (Fig. 3(b)), a BKT transition is determined with $T_{\rm{BKT}}$=4.35 K based on the relation $V \propto I^\alpha$ at $\alpha$=3 (Fig. 3(c) inset) \cite{halperin1979resistive,PhysRevB.80.214506}. It is worth noting that care has been taken to eliminate the impact of electron heating effects \cite{doron2020critical}, as discussed in the Supplementary Materials \cite{Supplement}.

The $T$-dependence of critical current density, $J_{\rm{c}}(T)$, provides insights into the SC nature (Fig. 3(c)). For a type II superconductor with thickness smaller than $\lambda (T)$, $J_{\rm{c}}(T)$ follows \cite{talantsev2015universal,zhang2023nodeless}:
\begin{equation}
J_{\rm{c}}(T) = \frac{H_{c1}(T)}{\lambda(T)}=\frac{\Phi_0}{4\pi\mu_0\lambda^3(T)}\mathrm{ln}(\kappa(T)+0.5) .
\label{J_cri}
\end{equation}
Here, $H_{c1}$ denotes the lower critical field, $\mu_0$ is the vacuum permeability, and $\kappa(T)$ is the Ginzburg–Landau parameter ($\kappa(T)=\lambda(T)/\xi(T)$). Since $\xi_{ab}(T)$ can be obtained from $H_{c2}^{c}(T)$, $\lambda(T)$ can be directly calculated from Eq. \ref{J_cri} without relying on any fittings (Fig. 3(d)). In Fig. 3(d), we compare both the $s$-wave and $d$-wave fittings to $\lambda(T)$ (see Ref. \cite{Supplement} for more details) \cite{tinkham2004introduction, talantsev2015universal}. It is evident that the $T$-dependence of $\lambda(T)$ matches well with the $s$-wave symmetry. The superconducting gap $\Delta (0)$ extracted from the fitting is $2.9k_{\rm{B}}T_c$ (1.1 meV), which is larger than the BCS weak coupling limit $1.76k_{\rm{B}}T_c$. Such a large gap aligns with the observed exceeding of the Pauli limit in $H_{c2}^{ab}(T)$. A more precise calculation of the Pauli limit can be derived from $H_P=\Delta/(\sqrt{g}\mu_B)$, which scales with the superconducting gap size, assuming no significant enhancement of the $g$ factor \cite{clogston1962upper}.

These findings taken together indicate the exfoliated CsV$_3$Sb$_5$ thick films have an anisotropic, fully gapped, $s$-wave pairing symmetry, which is similar to the bulk crystals, but exhibited distinct 2D SC, as revealed by the exceeding of Pauli limit for the in-plane upper critical field (Fig. 2(h)) and the $V$-$I$ characteristics analysis (Fig. 3(b)). Yet, the evidence of 2D SC is indirect. 

One of the smoking-gun evidence of 2D SC, as theoretically proposed, is a universal jump in the superfluid stiffness (also termed as superfluid density \cite{broun2007superfluid}), $\rho_s=\lambda^{-2}/\lambda_0^{-2}$, near the superconducting transition \cite{nelson1977universal,bishop1978study}. In Fig. 4(a), we show the $T$-dependence of $\rho_s$ for the 60-nm film and for the bulk \cite{gupta2022microscopic}, where an evident difference is identified. For the bulk crystal, $\rho_s$ is well described by the phenomenological $s$-wave model all the way up to almost $T_{\rm{c}}$ \cite{gupta2022microscopic}, while in the thin-film, agreement with $s$-wave description (the dashed curve in Fig. 4(a)) is limited to $T < 0.9T_{\rm{c}}$. As $T$ approaches $T_{\rm{c}}$, $\rho_s$ undergoes a much steeper, linear decrease (the dotted curve in Fig. 4(a)), which behaves exactly as the theoretical predictions \cite{nelson1977universal,bishop1978study} of the BKT transition. Meanwhile, we also include the universal transition relation $\pi\rho_s(T_{\rm{BKT}}) = 2T_{\rm{BKT}}$ (the black dashed line in Fig. 4(a)), which yields a transition at $T_{\rm{BKT}}$ = 4.24 K, consistent with the value obtained from Fig. 3(b).

\begin{figure}[ht]
\includegraphics[width=3.3in]{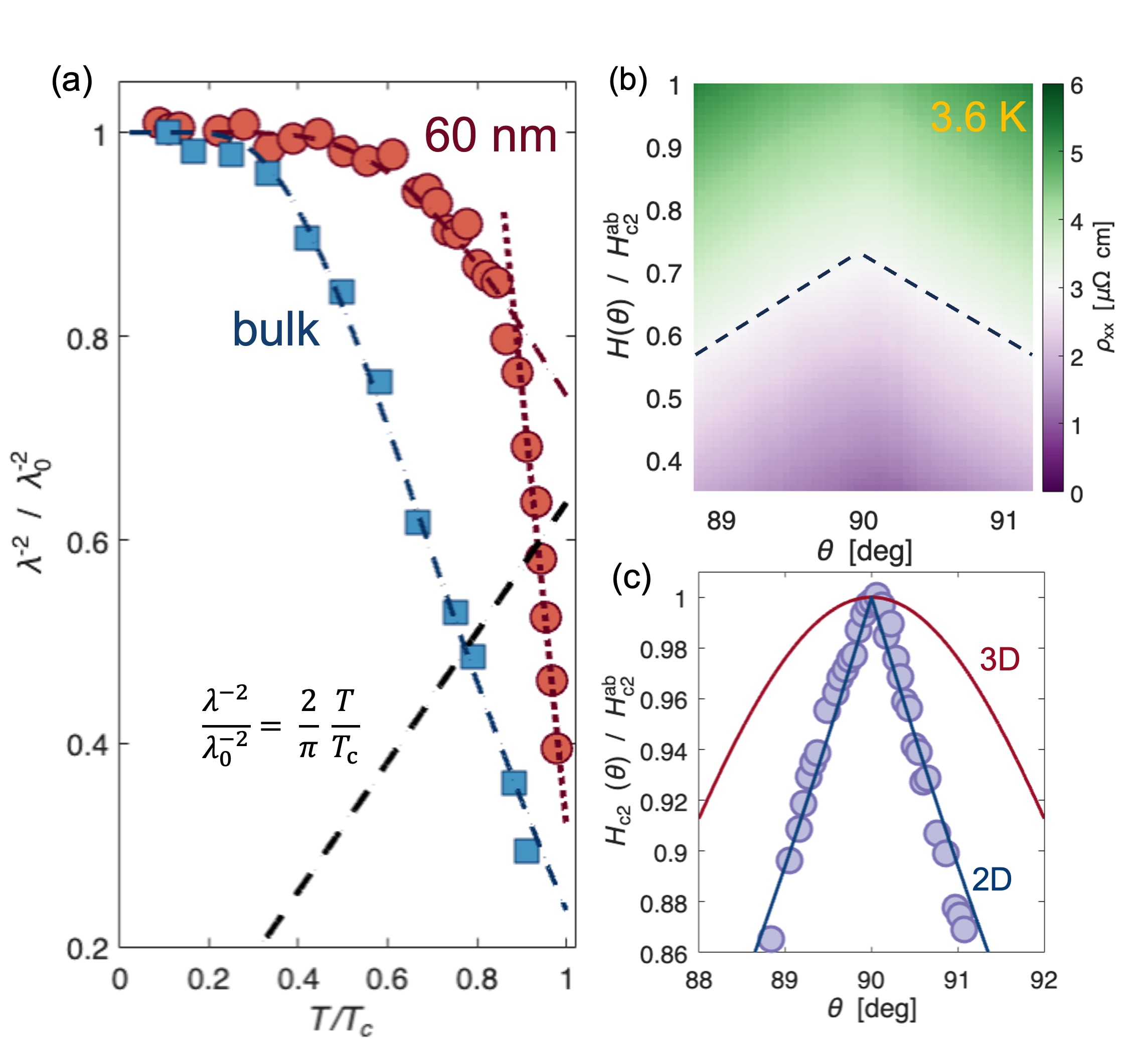}%  
\caption{\label{Fig4} Evidence of 2D SC. (a) The temperature dependent superfluid stiffness. The bulk data is from $\mu$SR measurement \cite{gupta2022microscopic}. The dashed red and blue curves are fits based on an $s$-wave order parameter. The experimental data of the 60-nm film drops faster than $s$-wave prediction near $T_{\rm{c}}$ in a linear form as indicated by the dotted red line. (b) $\rho_{xx} $ measured as a function of $H$ at fixed $\theta$, where $\theta$ is the angle between the applied field and the $c$-axis. The data were taken at 3.6 K. The magnetic field is normalized by $H_{c2}^{ab}$. The dashed curve is a guide to the eyes marking the upper critical field. (c) Angular dependent upper critical field. The solid curves are predictions from 3D Ginzburg-Landau model and 2D Tinkham model. }
\end{figure}

To further confirm the 2D SC characteristics, we conducted angular dependent measurements of $H_{c2}$, as shown in Fig. 4(b-c). We systematically tilt the magnetic field near 90 degrees (where the field is aligned along the in-plane direction). In Fig. 4(b), a colormap of the angular-dependent magnetoresistivity at a temperature near the transition, $\rho_{xx} (H, 3.6 \rm K)$, is shown where the dashed line is a guide to the eye to indicate the upper critical field $H_{c2}$. $H_{c2}$ (scaled by its maximum, $H_{c2}^{ab}$) is further shown in Fig. 4(c) as a function of the angle. A clear cusp-like peak for $H_{c2}$ is observed, which is a well-known feature to distinguish the 3D and 2D SC \cite{tinkham2004introduction}. In Fig. 4(c), the red and blue curves represent the predictions from 3D Ginzburg-Landau model $(H_{c2}(\theta) \sin \theta / H_{c2}^{ab} )^2 + ( H_{c2}(\theta) \cos \theta/ H_{c2}^c)^2 = 1$ and 2D Tinkham model $(H_{c2}(\theta) \sin \theta / H_{c2}^{ab} )^2 + \left| H_{c2}(\theta) \cos \theta/ H_{c2}^c \right| = 1$, respectively.  It is important to note that these curves are not fitting results; rather, they are derived from the experimentally measured parameters $H_{c2}^{c}$ and $H_{c2}^{ab}$.

The universal jump in superfluid stiffness near $T_c$ (Fig. 4(a)) and the cusp-shape peak of $H_{c2}$ (Fig. 4(c)) provide direct evidence of 2D SC in the 60-nm-thick exfoliated CsV$_3$Sb$_5$ film. Consistent results are also observed in a 40-nm-thick sample. This 2D superconducting state in the optimal thickness range forms in the clean limit, characterized by $\xi_{ab}/l_{ab} \approx$ 0.1, where $l_{ab}$ represents the electron mean free path and is estimated from the resistivity $\rho_{ab}$ (see Supplemental Material~\cite{Supplement}).  Such clean limit 2D superconductors provide opportunities for investigating finite momentum Cooper pairs, but they are rarely reported \cite{devarakonda2020clean}.

The observation of 2D SC in a thick exfoliated film with thickness around 40-60 nm is striking. The 60 nm thickness is approximately 70 times larger than the interlayer lattice spacing $d$, placing such thick film more akin to a 3D case rather than a 2D one. Conventionally, one straightforward way to realize 2D SC is to reduce sample thickness. As thickness decreases, the dimensional effect becomes increasingly significant, but such effect is only noticeable near the atomic limit, such as in atomic scale crystalline Pb films \cite{guo2004superconductivity}, monolayer or a-few-layer transition metal dichalcogenides \cite{de2018tuning}. The 60-nm-thick kagome film does not fall into this category.

Alternatively, 2D SC can be achieved by weakening interlayer coupling to effectively reduce the thickness of the superconducting layer. This approach allows for the existence of 2D SC in bulk crystals, independent of sample thickness. This is the case for TaS$_2$-related superlattices \cite{prober1977fluctuation, devarakonda2020clean,ribak2020chiral}, organic superconductors \cite{murata1985superconductivity} and cuprates \cite{Li2007two}. However, as indicated by Fig. 4(a), in contrast to the thick films, the bulk CsV$_3$Sb$_5$ does not exhibit 2D SC. Remarkably, such comparison suggests a crossover from 3D to 2D SC manifesting at thicknesses exceeding 60 nm.

Our findings position the thick CsV$_3$Sb$_5$ films as an unique case of 2D SC, which suggests potential for controlling 2D SC at the device level and exploring the 3D to 2D SC crossover beyond the atomic limit. The underlying mechanism remains to be explored, but it is worth highlighting that 2D SC occurs at an optimal thickness where the samples have both the highest $T_c$ and the lowest $T_{\rm{CDW}}$ \cite{song2023anomalous,Supplement}. This suggests that the coexisting CDW order might influence the superconducting properties, particularly by modifying the interlayer coupling. In cuprate superconductors, the mismatched CDW order is known to suppress interlayer coherence, leading to 2D SC and an enhanced $T_c$ \cite{Li2007two}. It would be interesting to investigate how the CDW order correlates with interlayer coupling along the $c$-axis in kagome systems. Our findings impose constraints in understanding the enhancement of SC at the optimal thickness.

To sum up, we have observed 2D SC in the thick exfoliated kagome films. The experimental evidence includes: 1) an abrupt jump in superfluid density near the transition; 2) a cusp-like feature in the angular dependent $H_{c2}$; 3) standard BKT analysis of the $V$-$I$ relations; 4) exceeding of the Pauli limit for the in-plane upper critical field. The observation of 2D SC in a thick film, akin to a 3D system, is a non-trivial finding, which suggests the presence of thin effective 2D superconducting layers and weak coupling between them. Our work sheds light on the hidden competing correlations within kagome superconductors as thickness decreases, which will spark interest of theoretical and experimental investigations regarding the crossover from 3D to 2D behavior.

\vspace{\baselineskip}
The authors thank A. P. Mackenzie, C. Hooley, J. G. Checkelsky, S. A. Chen and K. Semeniuk for useful discussions. FS and HZ are grateful to the Max Planck Society for financial support. SDW and ACS gratefully acknowledge support via the UC Santa Barbara NSF Quantum Foundry funded via the Q-AMASE-i program under award DMR-1906325.

\bibliography{CVS_2D_SC_Manuscript}% 

\end{document}

\preprint{APS/123-QED}

\title{Supplementary Materials for\\
\enquote{Two-dimensional superconductivity in a thick exfoliated kagome film}}

\author{Fei Sun}
\email{Correspondence: Fei.Sun@cpfs.mpg.de}
\affiliation{Max Planck Institute for Chemical Physics of Solids, 01187 Dresden, Germany}
\author{Andrea Capa Salinas}
\affiliation{Materials Department, University of California Santa Barbara,
Santa Barbara, California 93106, USA}
\author{Stephen D. Wilson}
\affiliation{Materials Department, University of California Santa Barbara,
Santa Barbara, California 93106, USA}
\author{Haijing Zhang}
\email{Correspondence: Haijing.Zhang@cpfs.mpg.de}
\affiliation{Max Planck Institute for Chemical Physics of Solids, 01187 Dresden, Germany}

\maketitle
\tableofcontents

\newpage
\section{Device preparation and transport measurements}

Exfoliated crystals were obtained by mechanically cleaving them from bulk crystals using a Scotch tape. Subsequently, they were transferred onto the silicon substrates capped with a layer of 90 nm thick silicon dioxide. The thicknesses of exfoliated crystals were measured by an atomic force microscope, and then crystals of desired thickness were selected for device fabrication. Electrical contacts were patterned by standard electron beam lithography, and Ti/Au electrodes were deposited by sputtering.  

A Quantum Design physical property measurement system (PPMS) He3 refrigerator,  equipped with a 9 T magnet, was used for transport measurements with externally connected electronics. SR830 lock-in amplifiers were used to perform four-point resistance measurements. The longitudinal resistivity was extracted by $\rho_{xx}= W t V_{xx}/(I L)$, where $W$ is the width, $L$ is the length, $t$ is the thickness of the device, $I$ is the alternating current applied from the source electrode and $V_{xx}$ is the measured voltage drop.

\section{Shubnikov-de Haas (SdH) quantum oscillations}

We have performed magnetoresistance measurements with a perpendicular magnetic field up to 9 T. The data taken at $T$=5 K is shown in Fig. S1. Clear SdH quantum oscillations can be seen at magnetic fields above 3 T. The magnetoresistance shows a monotonic increase with the applied magnetic field. The background trace was fitted using a polynomial function $\overline{\rho}_{xx} (H)= \sum_{i=0}^{4} \alpha_i H^i $. Fig. S1 shows the oscillatory components of the magnetoresisitance, which was extracted by subtracting the background trace:  $\Delta \rho_{xx} (H)$ = $\rho_{xx} (H) - \overline{\rho}_{xx} (H)$. The fast Fourier transform (FFT) for the oscillations reveals two beating frequencies at $F_1$=27 T and $F_2$=72 T, respectively. In Fig. S1 (c), we have highlighted the oscillations that correspond to the two frequencies listed in Fig. S1 (d).

\begin{figure*}[h]
    \centering   \includegraphics[width=0.7\linewidth]{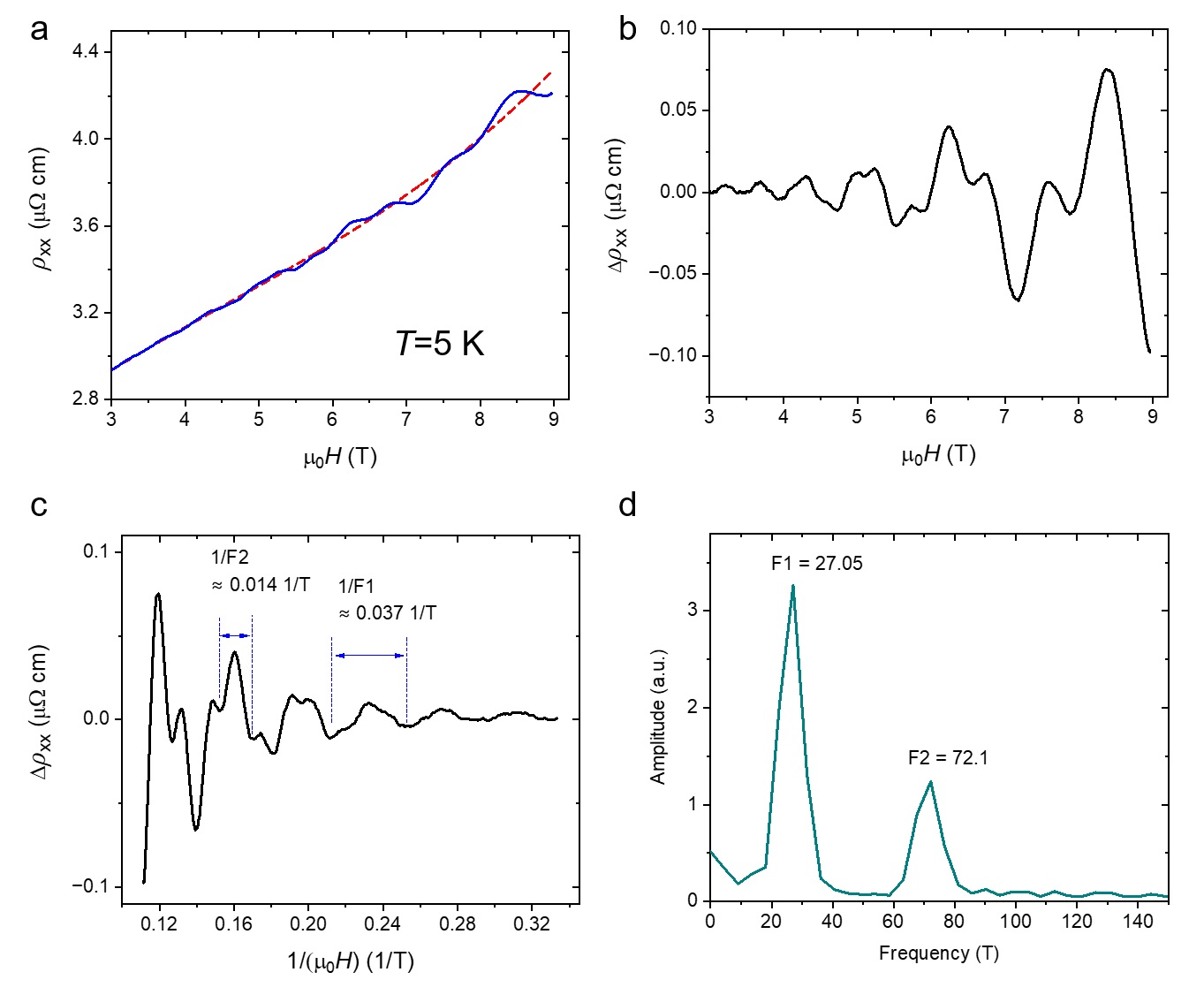}
    \caption{Quantum oscillations of the 60-nm-thick exfoliated flake. (a) Magnetoresistance $\rho_{xx} (H)$ is plotted as a function of perpendicular magnetic field taken at $T$=5 K. The dashed line is the background trace obtained by a fit to the data using a polynomial function $\overline{\rho}_{xx} (H)= \sum_{i=0}^{4} \alpha_i H^i $. (b) $\Delta \rho_{xx} (H)$ = $\rho_{xx} (H) - \overline{\rho}_{xx} (H)$ with the background trace subtracted. (c) $\Delta \rho_{xx} (H)$ is plotted as a function of 1/($\mu_0 H$). (d) Fast Fourier transform (FFT) for data shown in (c).  Two prominent frequencies can be obtained from FFT and they are associated with the periods labeled in (c). }
    \label{fig:FigS1}
\end{figure*}

\section{Estimates of the mean free paths}

As elaborated in the main text, the $ab$-plane mean free path can be estimated from the resistivity $\rho_{ab}$ in a two-dimensional case as \cite{mackenzie1998extremely}:
\begin{equation}
l_{ab} = \frac{hd}{e^2\rho_{ab}\Sigma_{i}k_{F}^i}
\label{MFP}
\end{equation}
where $h$ is the Planckian constant, $e$ is the elementary charge, $d$ is the interlayer spacing of 9 \AA, and $\Sigma_{i}k_{F}^{i}$ represents the sum of the Fermi wave vector $k_{F}$ at the Fermi surface. Using $\rho_{ab}$=2.2 \(\mu\Omega\cdot\text{cm}\) taken at temperatures around 5 K and a sum of $\Sigma_{i}k_{F}^{i}$=0.952 $\AA ^{-1}$ taken from Fermi surface mapping by photoemission spectroscopy \cite{PhysRevX.11.041030} is used in the calculation, $l_{ab}$ is estimated to be 110 nm at temperatures around 5 K.

The Fermi wave vector can also be calculated from the quantum oscillations, using the Onsager relation for 2D Fermi surfaces: $F=\frac{\hbar}{2 \pi e} S_F=\frac{\hbar}{2 \pi e} (\pi k_F^2) $. For frequencies of $F_1$=27 T and $F_2$= 72 T, we obtain $k_F$=0.029 and 0.047 $\AA ^{-1}$, respectively. The sum of the two $k_F$ is significantly smaller than the the $\Sigma_{i}k_{F}^{i}$ extracted from the photoemission spectroscopy, as the higher frequencies were not detected in the magnetic field range we applied. To avoid overestimating the mean free path, we were using the Fermi wave vectors read out from the photoemission spectroscopy.

The $c$-axis mean free path can be estimated from $l_c=l_{ab} \sqrt{\rho_{ab}/2\rho_{c}}$. From the known anisotropy ratio of $\rho_{c}$/$\rho_{ab}$ =600~\cite{ortiz2020cs}, $l_c$ is estimated to be 3.2 nm.

\section{Atomic force microscopy (AFM) image of the exfoliated flakes}

\begin{figure*}[h]
    \centering
    \includegraphics[width=0.55\linewidth]{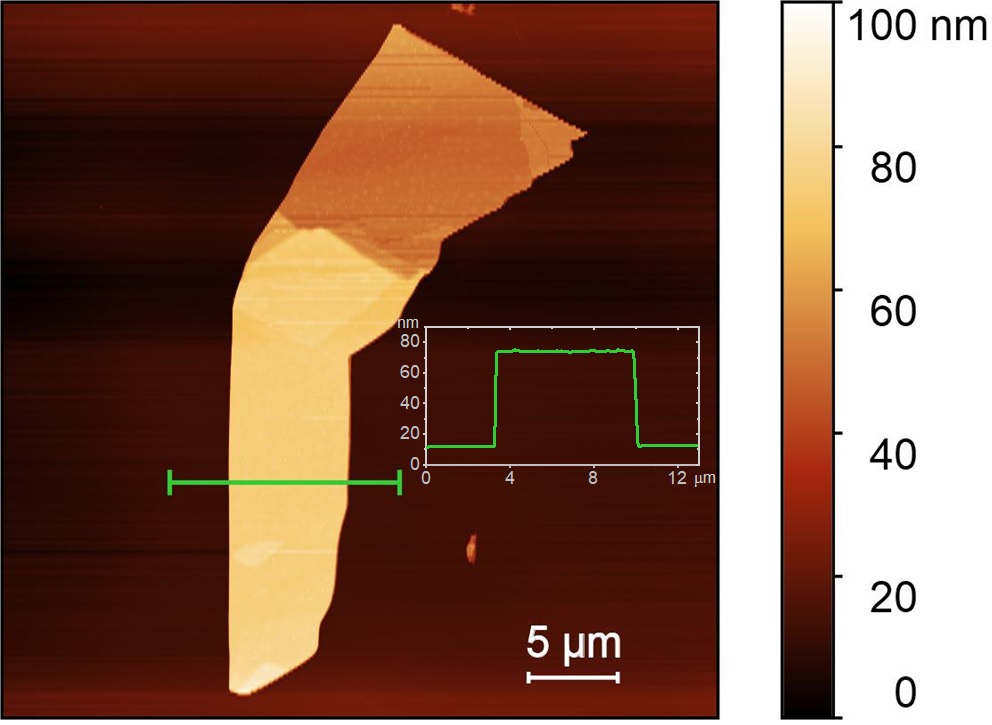}
    \caption{AFM image of one exfoliated flake. The thicknesses of the exfoliated flakes are measured from the step height as shown in the inset.}
    \label{fig:FigS2}
\end{figure*}

To precisely measure the thicknesses of our exfoliated flakes, we conducted the AFM scanning. Fig. S2 shows the AFM image of one exfoliated flake and the thickness is extracted from the step height as shown in the inset.

\newpage
\section{The temperature dependence of the upper critical field anisotropy}

The anisotropy of the superconducting states is characterized by the anisotropic ratio, defined as $H_{c2}^{ab}/ H_{c2}^{c}$, which is associated with the Ginzburg-Landau coherence lengths by the relation $H_{c2}^{ab}/ H_{c2}^{c}$ = $\xi_{c}/ \xi_{ab}$. Using the WHH model, the anisotropy can be extracted from the slope of the upper critical fields $dH_{c2}/dT$ around $T_c$, as this slope manifests the orbital limit.  Here, in Fig. S3, we plot the deduced anisotropic ratio as a function of $T$. As $T$ approaches $T_c$, the anisotropic ratio approaches a value of 40. The ratio decreases down to 9 at low temperatures. The derived anisotropic values are significantly higher than those reported in iron-based superconductors \cite{jaroszynski2008upper}, comparable to those in cuprates, and smaller than transition metal dichalcogenides whose superconductivity was realized by quantum confinement \cite{lu2015evidence}.

\begin{figure*}[!htbp]
    \centering
    \includegraphics[width=0.6\linewidth]{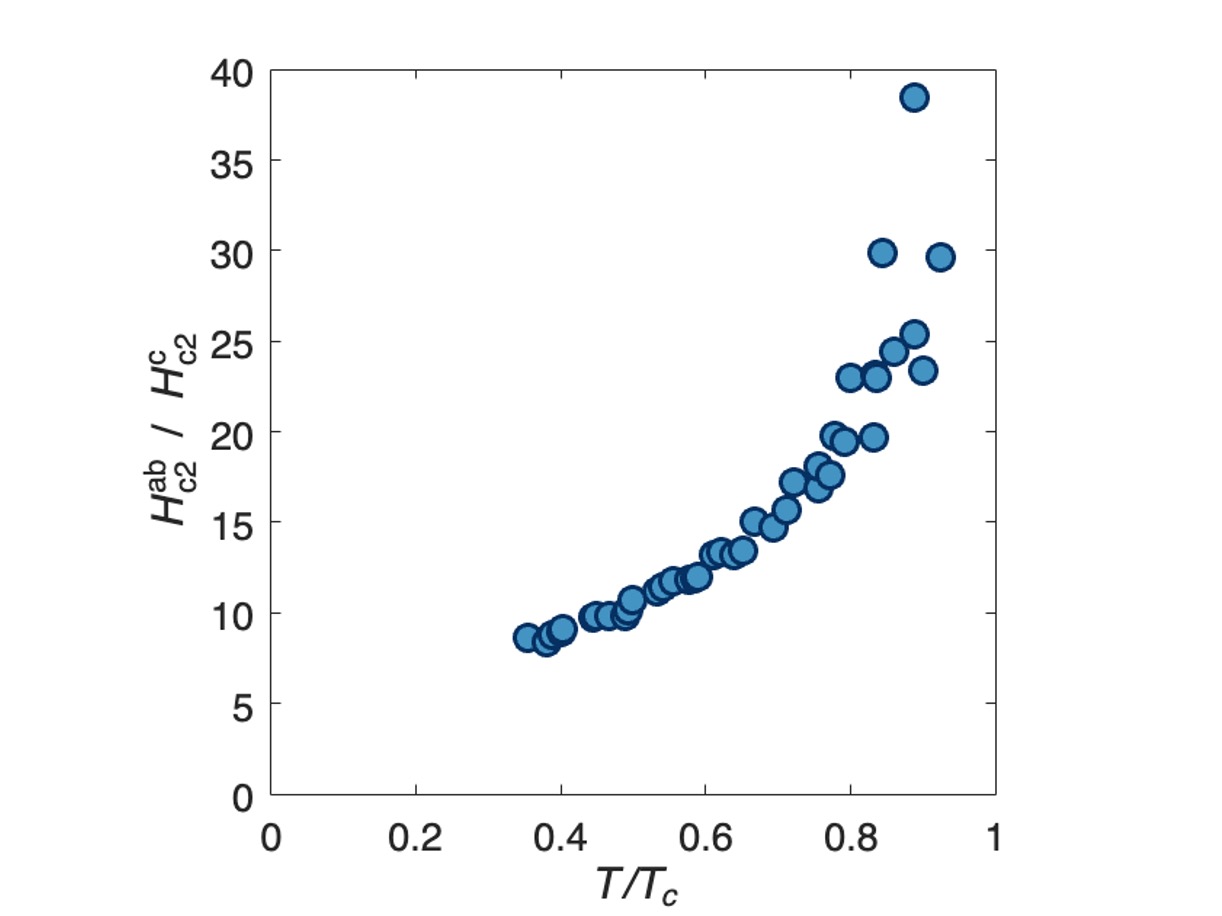}
    \caption{Temperature-dependent upper critical field anisotropy.}
    \label{fig:FigS3}
\end{figure*}

\newpage
\section{Anisotropy of the orbital limit}

As discussed in the main text, the orbital limit of the superconducting states can be represented by the slope of the upper critical fields near the transition temperature. To obtain the anisotropy of the orbital limit, we plot both $H_{c2}^{ab}$ and $H_{c2}^{c}$ as a function of the reduced temperature $T/T_c$ near $T_c$ in Fig. S4. For the case of an in-plane field, the slope $dH_{c2}^{ab}/d(T/T_c)$ is well defined to be around -16.3 T, while for the out-of-plane field, the slope $dH_{c2}^{ab}/d(T/T_c)$ ranges from -0.72 to -0.86 T. Therefore, the anisotropy ratio is $dH_{c2}^{ab}/d(T/T_c)$/$dH_{c2}^{c}/d(T/T_c)$ is around $18 \sim 22$, which is twice as large as that in the bulk \cite{ni2021anisotropic}.

\begin{figure*}[h]
    \centering
    \includegraphics[width=0.6\linewidth]{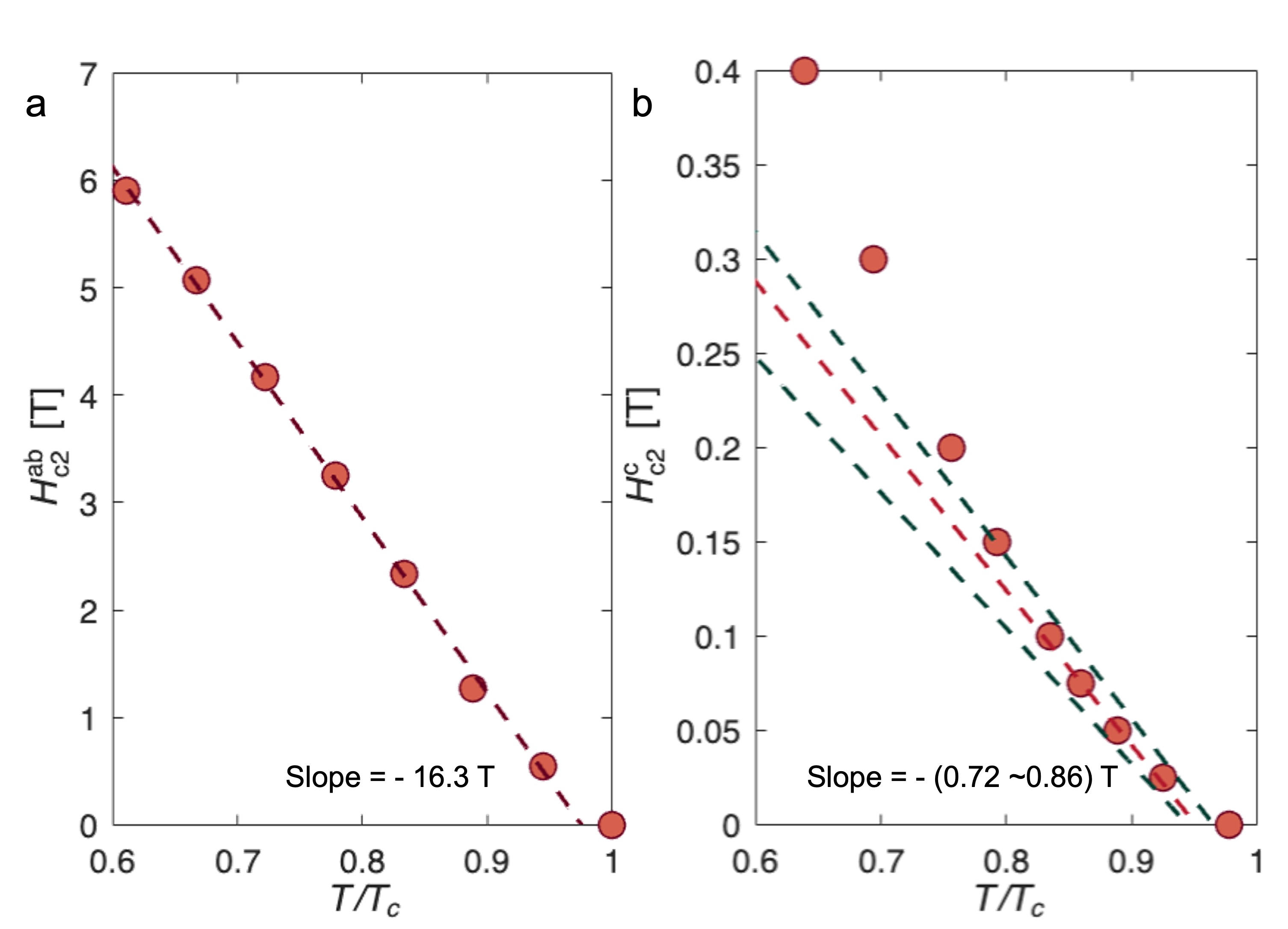}
    \caption{The orbital limit of the upper critical fields for both directions of the magnetic field. The dashed lines are the linear fittings, while the two black dashed lines in (b) mark the upper and lower boundary of the slope.}
    \label{fig:FigS4}
\end{figure*}

\section{A comparison of $H_{c2}$ between the films and the bulk}

To explore the 2D superconducting nature, an effective avenue is to explore the thickness variation, as illustrated in Fig. S5. The concave feature in $H_{c2}^{c}$ (Fig. S5(a)) is more prominent as the thickness reduces from bulk to 60 nm, with $H_{c2}^{c} (0)$ reaching its peak for the 60-nm-thick sample. Additionally, our measurements on a 40-nm-thick sample did not show a further enhancement of $H_{c2}^{c}$ either (See Fig. S7 below)). Furthermore, a more distinct difference between the film and the bulk is observed in the in-plane upper critical field $H_{c2}^{ab}$, as depicted in Fig. S5(b). Comparing the $H_{c2}$ for different orientations, near $T_{\rm{c}}$, the in-plane upper critical fields (Fig. S5(b)) exhibit a more distinct difference between bulk and film than the out-of-plane direction (Fig. S5(a)). Quantitatively, the anisotropy of the orbital limit, represented by the slope of $H_{c2}$ near the transition ($dH_{c2}^{ab}/dT)|_{T_c}/(dH_{c2}^{c}/dT)|_{T_c}$ \cite{PhysRev.147.295}, in the 60-nm film is twice as large as that in the bulk.

\begin{figure*}[h]
    \centering    \includegraphics[width=0.6\linewidth]{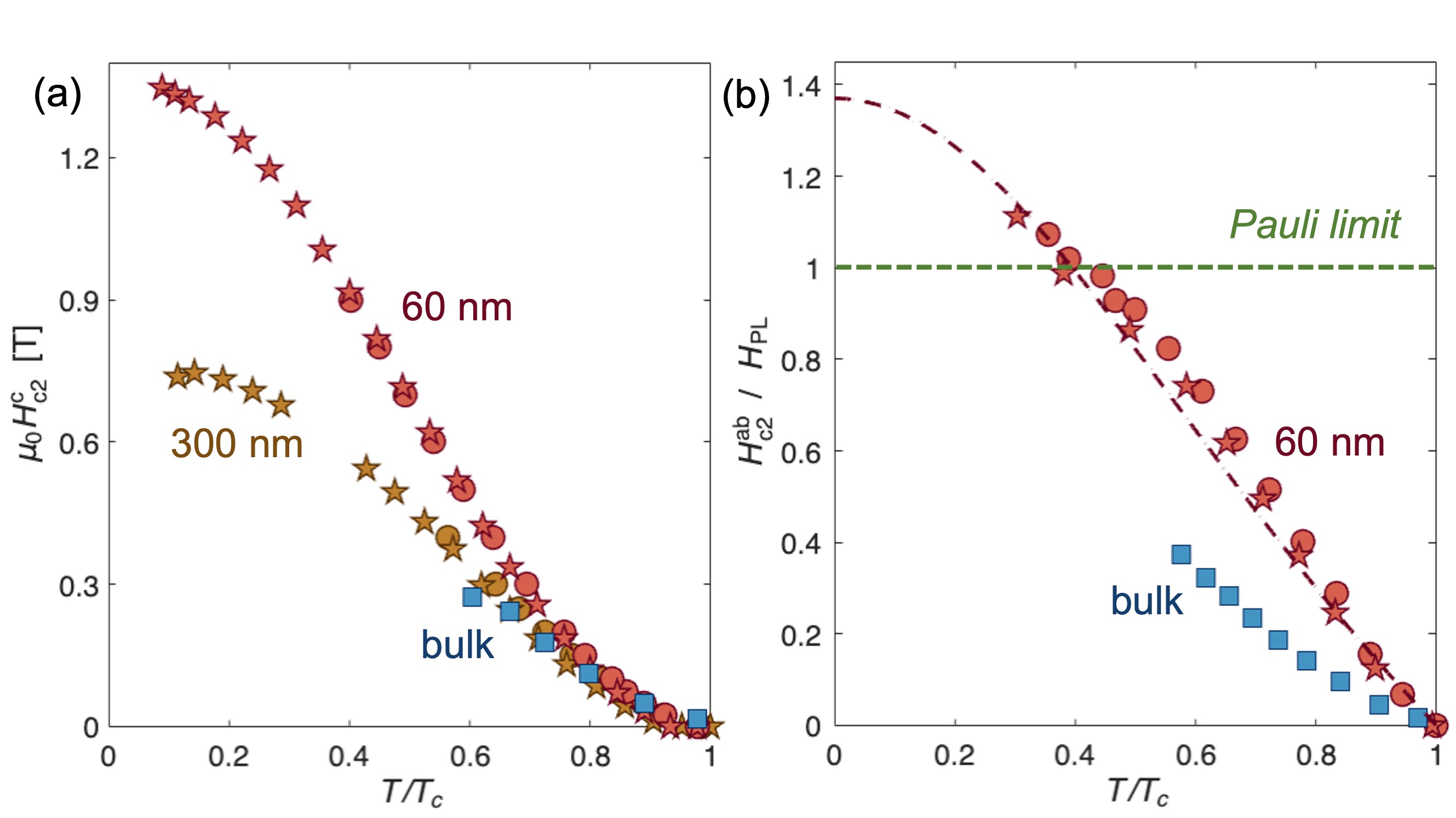}
    \caption{The comparison between the films and the bulk. (a) and (b) are the comparison of $H_{c2}^{c}$ and $H_{c2}^{ab}$, respectively. The stars and the circles represent $H_{c2}$ extracted from the temperature- and field- sweeps, respectively. The bulk data is adopted from Ref. \cite{ni2021anisotropic}. The dashed curve in (b) is a guide curve to the eyes based on the Ginzburg-Landau theory \cite{tinkham2004introduction}.}
    \label{fig:FigS5}
\end{figure*}

\section{The temperature dependence of the penetration depth}

As discussed in the main text (Fig. 3(d)), the temperature dependent penetration depth offers information of the superconducting pairing symmetry and gap size. Here we show the fitting function of $\lambda$ for different superconducting pairing symmetries \cite{tinkham2004introduction, talantsev2015universal}. For $s$-wave symmetry:

\begin{equation}
\frac{\lambda^{-2}(T)}{\lambda^{-2}(0)} = 1-2\sqrt{\frac{\pi\Delta(0)}{k_BT}}e^{-\frac{\Delta(0)}{k_BT}} 
\label{swave}
\end{equation}

and for $d$-wave symmetry:

\begin{equation}
\frac{\lambda^{-2}(T)}{\lambda^{-2}(0)} = 1-\sqrt{2}\frac{k_BT}{\Delta_m(0)} 
\label{dwave}
\end{equation}

where $\Delta (0)$ is the $s$-wave superconducting gap and $\Delta_m(0)$ is the maximum value of the $d$-wave superconducting gap.

\newpage
\section{$H_{c2}$ vs $T$ data of the 40-nm-thick exfoliated crystal}

A 40-nm-thick exfoliated CsV$_3$Sb$_5$ crystal has been measured, and the upper critical fields along different directions are plotted in Fig. S6. For the magnetic fields applied along the $c$-axis, $H_{c2}^c(T)$ shows a similar concave feature and, at the low temperature limit, $H_{c2}^c$ saturates at around 1.2 T, which is slightly lower than that of the 60-nm-thick sample. For the in-plane upper critical fields, $H_{c2}^{ab}$ exceeds the Pauli paramagnetic limit $H_P=1.84$ $T_c $ [T/K] at temperatures below 2 K .

\begin{figure*}[h]
    \centering \includegraphics[width=0.55\linewidth]{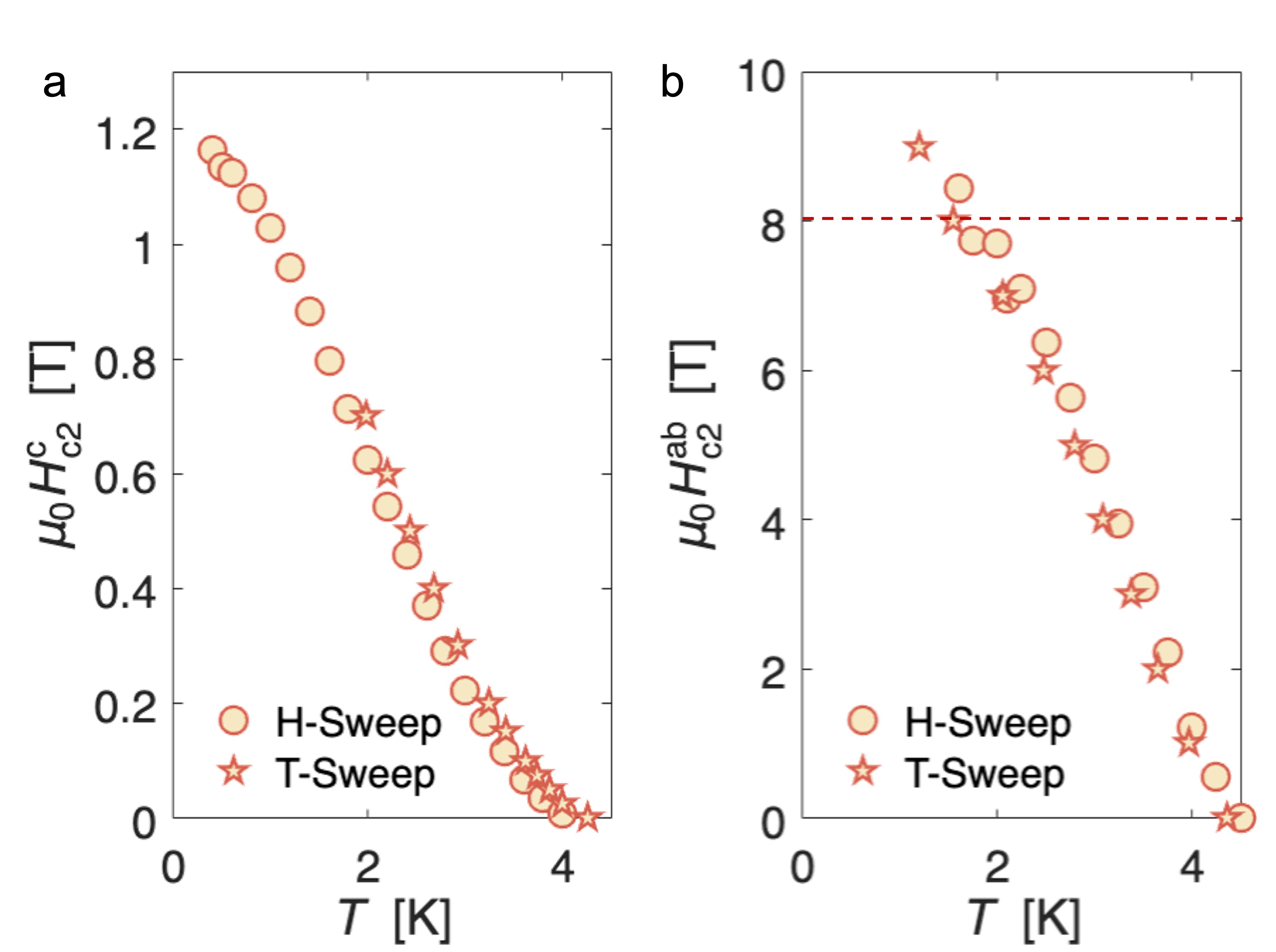}
    \caption{The $H_{c2}$ - $T$ phase diagrams of the 40-nm-thick sample.}
    \label{fig:FigS6}
\end{figure*}

\section{Thickness dependence of the superconducting and CDW transitions}

We have investigated the thickness dependence of the superconducting transition temperature, $T_{\text{c}}$, and the CDW transition temperature, $T_{\text{CDW}}$. The critical temperatures are extracted from the temperature dependent resistance measurements, and the results are summarized in Fig. S7. Consistent with previous reports~\cite{song2021competition,song2023anomalous}, with the thickness reducing from bulk to 40 nm, the $T_{\text{CDW}}$ is decreasing from 94 K to 76 K, and $T_{\text{c}}$ is enhanced from 2.5 K to 4.5 K. 

\begin{figure*}[h]
    \centering
    \includegraphics[width=0.45\linewidth]{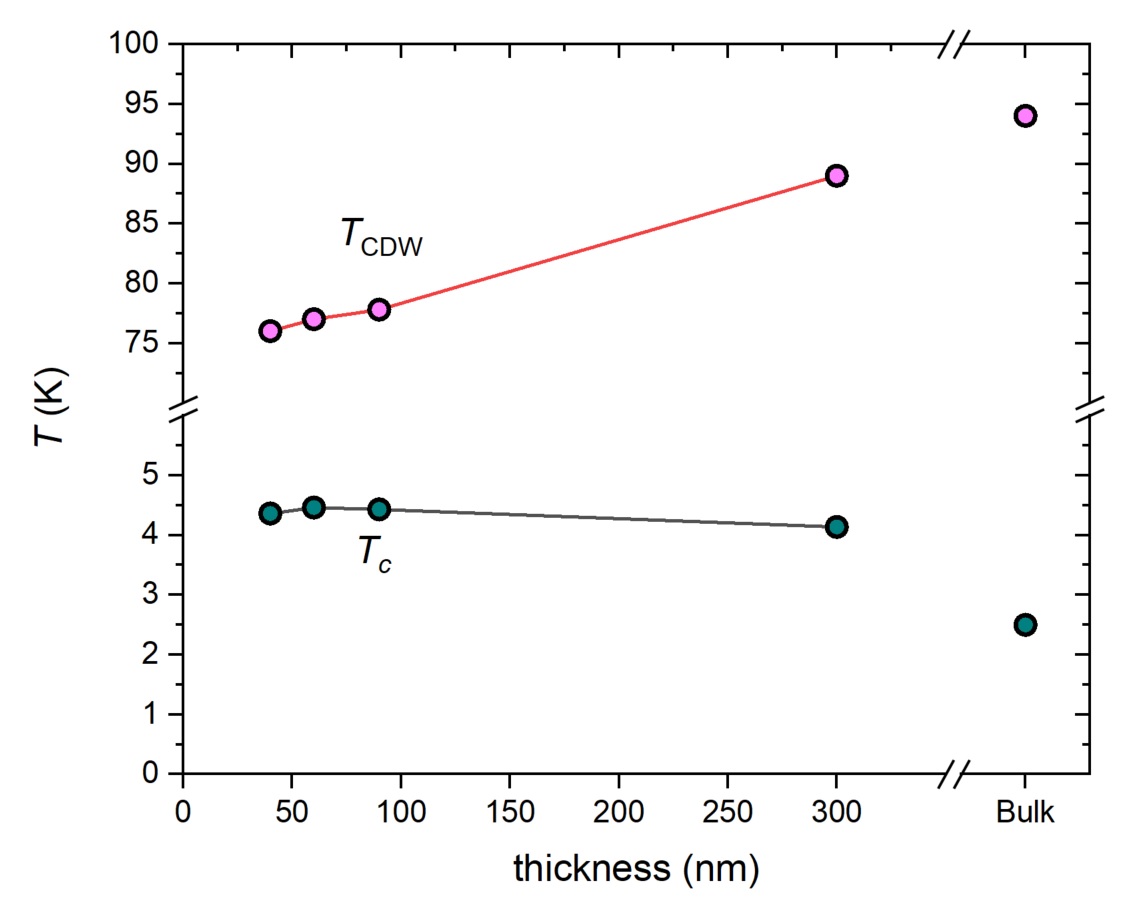}
    \caption{Thickness dependence of $T_{\text{c}}$ and $T_{\text{CDW}}$. The data for the bulk is taken from~\cite{ortiz2020cs}.}
    \label{fig:FigS7}
\end{figure*}

\section{The effect of electron heating in $I-V$ characteristics}

Electron heating can be detrimental to the superconductivity measurements. To eliminate the possibility of electron overheating, we implemented various measurement schemes when taking $I-V$ measurements.

We conducted two types of measurements to evaluate the heating effect \cite{weitzel2023sharpness}: fast measurements (lasting 6 to 24 seconds per sweep) and slow measurements (10 minutes per sweep). The fast measurements were conducted in pulse mode, with a pulse width of 20 ms for each data point. The slow measurements were performed using the standard DC mode, with a delay time of 500 ms at each data point. The results, plotted in Fig. S8, show that data from both fast and slow measurements overlap, with no significant differences above the background noise level (indicated by the grey shaded region). This demonstrates that electron heating is negligible in our devices.

We then extracted the power law exponents $\alpha$ based on the relation $V \propto I^\alpha$ in the lowest possible regime (above the background noise level). The BKT temperature is determined to be $T_{\text{BKT}}$=4.15 K for the 40-nm-thick film.

\begin{figure*}[h]
    \centering
    \includegraphics[width=0.4\linewidth]{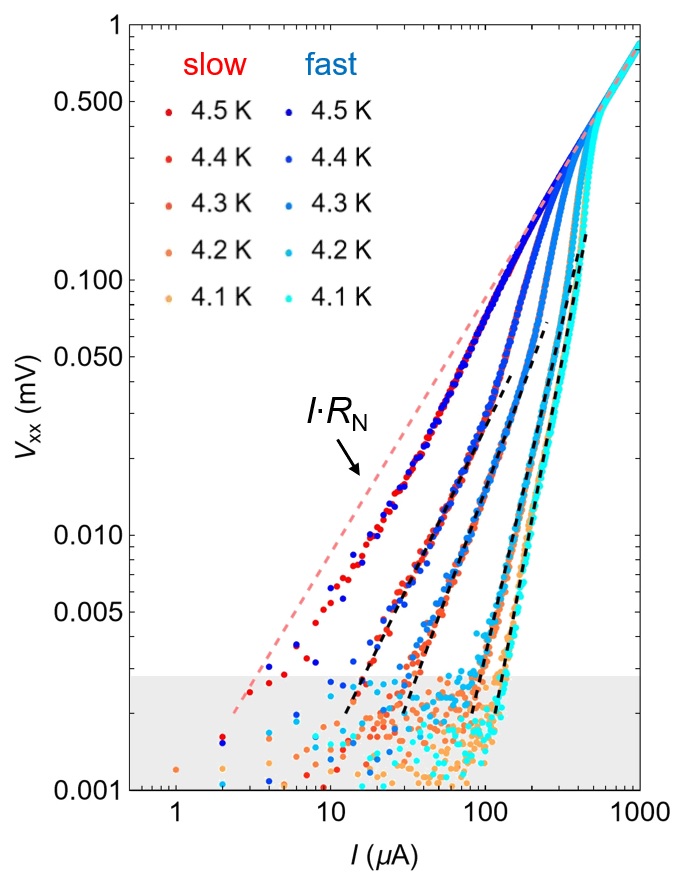}
    \caption{The $V$ vs $I$ curves of a 40-nm-thick film taken near $T_{\text{BKT}}$. The pink dashed line represents the normal state: $V(I)=I \cdot R_N$.   The black dashed lines are the fittings based on $V \propto I^\alpha$, with $\alpha$=1.24, 1.65, 2.625, 3.2 (from left to right), respectively. The grey shaded region denotes the background noise level.}
    \label{fig:FigS8}
\end{figure*}

\newpage

\section{Further analysis of the electron heating effect in clear-limit system}

As suggested in Ref. \cite{doron2020critical}, the electron over heating can contribute to the non-linear behavior of $V$-$I$ curves, especially in the disordered systems \cite{postolova2015nonequilibrium, weitzel2023sharpness}. Here in this section, we demonstrate in our case which is a clean limit superconducting, the electron over heating effect is much weaker.
The electron over heating effect occurs when the power dissipated by the measurement current exceeds the rate of heat removal from the electrons. This is governed by the heat balance equation:

\begin{equation}
P = \Gamma\Omega(T^{\beta}\rm_{el}-T^{\beta}\rm_{ph})
\label{HeatingEffect}
\end{equation}

Here, $\Omega$ is the volume of the sample, $\Gamma$ and $\beta$ are the the parameters characterizing the process of the heat transferring from the electron system to the phonon bath via electron-phonon coupling, $T\rm_{el}$ and $T\rm_{ph}$ are the temperatures of the electron and phonon sub-system. $P$ is the heating power, which can also be expressed as $P=IV$ or $P=I^2 R(T\rm_{el} (\it I))$. For a given current $I$, the heating power depends linearly on the resistance. When the current is large enough (or when the heating effect is the dominant contribution), one has $T\rm_{el} \gg T\rm_{ph}$. Therefore, ignoring the second term on the right-hand-side and taking the logarithm of both sides of the Eq. S4, one has:

\begin{equation}
\rm log \it P = \beta \rm log (T\rm_{el})+log(\Gamma\Omega))
\label{HeatingEffect2}
\end{equation}

Therefore, $\rm log \it P$ goes linearly with $\rm log (T\rm_{el})$ in this case. A typical example can be find in Ref. \cite{doron2020critical} where for sufficient high current, the $\rm log \it P$ - $\rm log (T\rm_{el})$ for multiple temperatures all collapse onto the line described by Eq. S5. However as we show in Fig. S9, in our case, this collapsing behavior does not happen even under the highest current applied in our experiments, which implies that the electron over heating effect is not dominating in the $I-V$ measurements.

\begin{figure*}[h]
    \centering
    \includegraphics[width=0.6\linewidth]{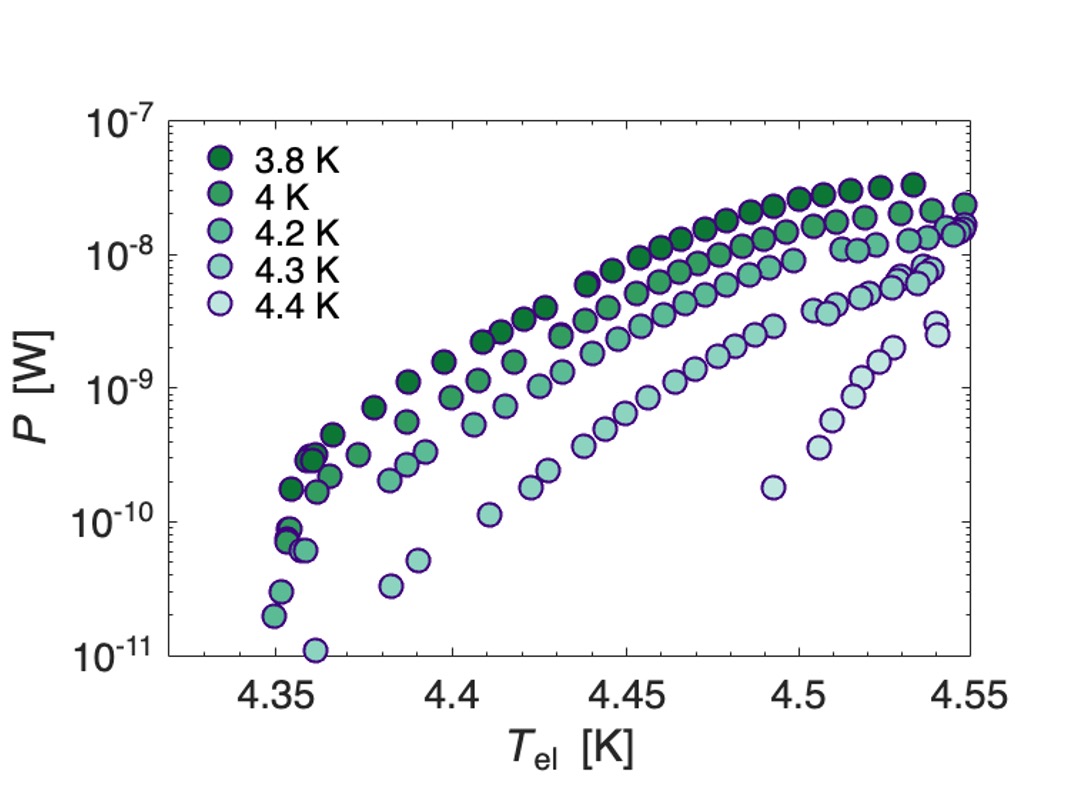}
    \caption{Electron overheating analysis based on Eq. S5.}
    \label{fig:FigS9}
\end{figure*}

\newpage
\bibliography{CVS_2D_SC}